\documentclass{elsart3p}

\usepackage{natbib}

 \usepackage{graphicx}

\usepackage{amssymb}

\usepackage{lineno}

\begin{document}

\begin{frontmatter}



\title{First results from the NEMO Phase 1 experiment}

\author[label1,label2]{Isabella Amore}  {for the NEMO Collaboration}
\address[label1]{Dipartimento di Fisica e Astronomia, Universit\`{a} di Catania, Italy}
\address[label2]{INFN Laboratori Nazionali del Sud, Catania, Italy}
\ead{amore@lns.infn.it}

\begin{abstract}
The NEMO prototype detector, called ``NEMO Phase-1", has been successfully operated at 2000 m depth from December 2006 to May 2007. The apparatus comprises a \emph{Junction Box} and a \emph{Mini-Tower} hosting 16 optical sensors. Preliminary results are presented. Positions of the optical sensors in the Mini-Tower were reconstructed through the acoustic positioning system with high level accuracy. Environmental parameters were analyzed.
From data corresponding to a livetime of 11.3 hours, atmospheric muon tracks have been reconstructed and their angular distributions were measured and compared with Monte Carlo simulations.
\end{abstract}


\end{frontmatter}

\section{NEMO Phase-1}
\label{sec:nemo}
The NEMO Phase-1 experiment is a technological demonstrator for a km$^{3}$ underwater neutrino detector. Its final aim is to validate key technologies (mechanics, electronics, data transmission, power distribution, acoustic positioning and time calibration systems) proposed for the km$^{3}$ detector.
The apparatus includes prototypes of the critical elements of the km$^{3}$ detector: the \emph{Junction Box} (JB) and a prototype detection unit called \emph{Mini-Tower}. On $18^{th}$ December 2006 it was successfully installed 25 km East off-shore Catania at 2080 m depth at the underwater Test-Site of the Laboratori Nazionali del Sud in Catania.
The Junction Box was connected to shore via a 28 km electro-optical cable reaching a shore station located inside the port of Catania where the power supply, the instrumentation control and the data acquisition systems are located. The Mini-Tower and the Junction Box were interconnected with an electro-optical jumper equipped with wet mateable hybrid connectors.
Connection operations were performed with an underwater Remotely Operated Vehicle (ROV). The detector was turned off on 18$^{th}$ May 2007. For more details see the references \cite{Migneco08,Capone08}.

In this work, preliminary results from the NEMO Phase-1 detector are presented. Data from the acoustic positioning system and environmental parameters were analyzed.
Atmospheric muon tracks have been reconstructed and results were compared with Monte Carlo simulations.

\section{The Mini-Tower}
The prototype detection unit built for NEMO Phase-1 is a \emph{``Mini-Tower"} \cite{Musumeci06} of 4 floors; each floor is made of a 15 m long structure hosting two optical modules (one down-looking and one horizontal-looking) at each end (4 OM
per floor). The floors are vertically spaced by 40 m. Each floor is connected to the following one
by means of four ropes that are fastened in a way that forces each floor to take an orientation perpendicular
with respect to the adjacent (top and bottom) ones. An additional spacing of 80 m is added between the tower base and the lowermost floor to allow for a sufficient water volume below the detector.

In addition to the 16 Optical Modules (OMs) the instrumentation installed on the Mini-Tower includes several sensors for calibration and environmental monitoring.
Two hydrophones (labeled H0 and H1) are installed at the end-points of each tower-floor, close to the optical modules, and two additional monitor hydrophones (called ``monitoring station") are installed on the tower base.
The hydrophones, together with an acoustic beacon placed on the tower base, and other three beacons installed on the sea bed, are used for precise determination of the tower position by means of time delay measurements of acoustic signals. The other environmental instruments are: a Conductivity-Temperature-Depth (CTD) probe used for monitoring the seawater temperature and salinity, a light attenuation meter (C-Star) and an Acoustic Doppler Current Profiler (ADCP) that provides continuous monitoring of the deep sea currents along the whole tower
height. In addition, a compass-tiltmeter board is placed on each floor in order to measure the tower floor inclination and orientation.

\section{Acoustic Positioning System}

A mandatory requirement for the muon tracking is the knowledge of the absolute position of the optical sensors.
In fact, while the position and orientation of the tower base is fixed and known from its installation, the rest of the structure can bend under the influence of the sea currents.
A precise determination of the PMTs position is obtained with a system based on
time of flight measurements of acoustic pulses between a fixed sea-floor array of acoustic beacons (that form a Long Base Line, LBL) and the target hydrophones located on the detector structure. Hydrophone positions are deduced applying a triangulation method.
Measurements were performed at regular intervals of ~$\sim$1 second. Time of flight (TOF) is defined as the difference between the time of arrival (TOA) of the acoustic signal on the hydrophone and the time of emission on the beacon (TOE).

\begin{figure}
\begin{center}
  \includegraphics[width=8.5cm]{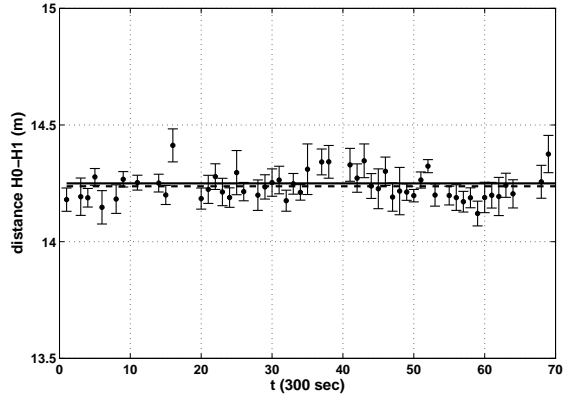}\\
\end{center}
  \caption{Distance H0-H1 measured for floor 2. Each point represents the average distance over a period of 300 s. For comparison the distance measured during the integration phase ($d=14.25 \pm 0.01$ m, solid line) is shown.}
  \label{fig:accuracy}
\end{figure}

To recognize the beacon pulses a technique called Time Spectral Spread Codes (TSSC) was adopted in this application.
Each beacon transmits a pattern of 6 pseudo-random pulses (spaced by ~$\sim$1 sec) that is different from the others. Each pulse length was 5 ms and the sequence of pulses was built in such a way to avoid overlap between two consecutive pulses. In this way a typical beacon pulse sequence can be recognized without ambiguity and all the beacons can transmit their characteristic pulse sequence at the same acoustic frequency. This represents a real advantage since all beacons can be identical except for the software configuration that defines the pulse sequence, and receivers can be sensitive to only one acoustic channel.

In order to obtain the required accuracy of 15 cm (comparable with the dimensions of the PMT)
the time of flight must be evaluated with an accuracy of the order of 10$^{-4}$ sec.
To achieve this goal an accurate calibration of the LBL must be
performed and the clock drift of the stand-alone beacons must be taken into account.
The beacon absolute positions and relative distances were determined, acoustically, at the time of detector installation, using a ROV equipped with a 32 kHz pinger, GPS time synchronized, and a high accuracy pressure sensor.
Once the distances between the beacons and the monitoring station, and the sound velocity have been determined, one is able calculate the TOF. The TOE of the beacon pulse, in the Master Clock reference time, is obtained measuring the TOA of this pulse at the monitoring station. This procedure allows also to compensate the clock drift of the stand-alone beacons (about 20 ns/s) during the whole live time of the apparatus.
Further, in order to be able to merge in post-processing positioning data together with optical modules detection information, both were time stamped with a universal time reference tag (the Universal Time Coordinate, UTC).

The relative distances between beacons were measured with an accuracy of ~$\sim10$~cm. In order to estimate the accuracy of the positioning system, distances between hydrophones H0 and H1 on the floor were measured. In Fig.\ref{fig:accuracy}, as an example, the distance H0-H1 measured for floor 2 is shown. Each point represents the average distance measured over a period of 300 s, in the time interval from 1$^{st}$ February h.17 to 1$^{st}$ February h.23 (6 hours). The mean value of the measured distance is 14.24 $\pm$ 0.06 m (dashed line). This value is compared with the distance H0-H1, equal to 14.25 $\pm$ 0.01 m (solid line), measured on-shore, during tower integration. This result indicates that the obtained accuracy in the determination of hydrophones positions is less than 10 cm, a value better than the design requirement (15 cm).

Data of acoustic positioning system, for the full period of the detector operation (from $18^{th}$ December 2006 to $18^{th}$ May 2007), were extensively analyzed. Tower positions were reconstructed and movements, as a function of time, were studied within a long and short time scale \cite{Amore08}.

\section{Underwater sea currents}

The Acoustic Doppler Current Profiler (ADCP) provides continuous monitoring of the deep sea currents. This instrument, mounted on the $4^{th}$ floor, downward oriented, measures the underwater currents below that depth, in a range of about 150 m. The ADCP data in the period of the detector data taking were analyzed.

\begin{figure}
\begin{center}
  \includegraphics[width=8cm]{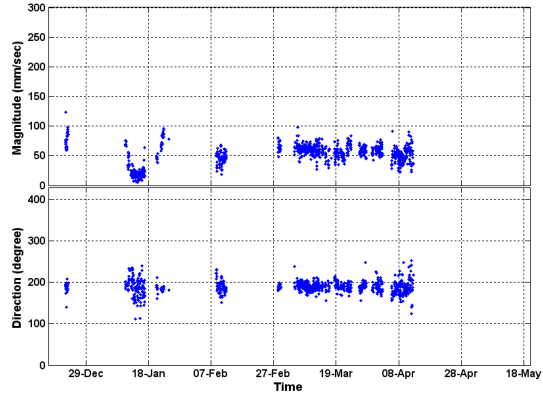}\\
\end{center}
  \caption{Magnitude and direction of the underwater sea currents, as a function of time, measured by the ADCP.}
  \label{fig:adcp}
\end{figure}

In Fig.\ref{fig:adcp} the magnitude and direction of the sea current, measured with the ADCP, are shown. In this plot are reported the values averaged in 5 bin of depths, corresponding to depth values between 12.5 and 25 m below floor 4. The recorded current magnitude is always less than 10 cm/sec and the average direction is $\sim180^{\circ}$, this means that the current is directed towards the South.

\section{PMTs background counting rate}
Each of the 16 OM hosts a 10'' PMT Hamamatsu R7081-Sel.
The instantaneous rate value is calculated by the Front-End board of the PMT averaging, in a time window of 1 $\mu$s, all the hits whose amplitude exceeds a given threshold equivalent to 0.3 (spe).
Background hit rates, for the full period of the detector operation, were analyzed. The average rate of photon hits on all PMTs was measured to be $\approx 80$ kHz as expected from $^{40}$K
decay plus a contribution due to diffuse bioluminescence.
The fraction of time in which a PMT records a high photon rate, due to bioluminescence bursts, is typically of the order of few percent, which implies a negligible dead time for the detector in
the search for muon events.

\begin{figure}
\begin{center}
  \includegraphics[width=7.5cm]{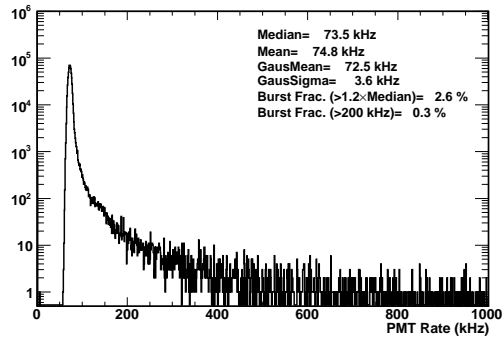}\\
\end{center}
  \caption{Histogram of photon hit rate distribution for a PMT located on the 4$^{th}$ floor, recorded on $10^{th}$-$20^{th}$ January 2007.}
  \label{fig:rate}
\end{figure}

\begin{figure}
\begin{center}
  \includegraphics[width=8cm]{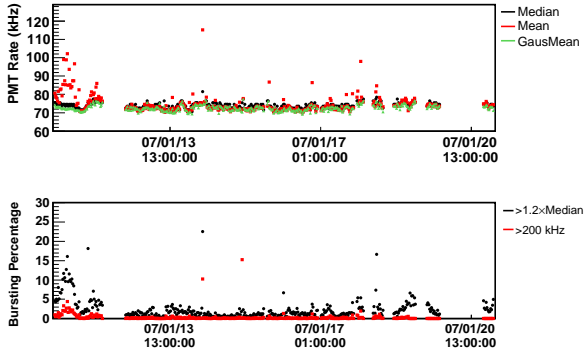}\\
\end{center}
  \caption{Top panel: PMT baseline rate as a function of time, evaluated in term of the median, mean of photo hit rate distribution and as average value of the Gaussian fit of the histogram peak (see Fig.\ref{fig:rate}). Bottom panel: burst fraction evaluated as the percentage of time in which the rate is larger than 1.2 times the median rate value or larger than 200 kHz.}
  \label{fig:rate_time}
\end{figure}

In Fig.\ref{fig:rate}, as an example, the histogram of the rate distribution is shown for a PMT located on 4$^{th}$ floor, in the time interval between $10^{th}$-$20^{th}$ January 2007.
In order to disentangle the slowly variable contribution to optical background due to $^{40}$K decay and diffuse bioluminescence, the so called \emph{baseline}, defined as the median value of the rate distribution, was calculated.
The baseline of the PMT rate was 73.5 kHz. For comparison the mean value of the distribution and the average value of the Gaussian function fitting the peak are reported.
This distribution shows also a tail extended to several hundreds kHz due to intense bioluminescence bursts.
To evaluate the contribution of the bioluminescence bursts, the so called \emph{burst fraction}, defined as the percentage of time for which the rate exceeds a given threshold, was calculated.
Burst fraction was calculated in two different ways: as the percentage of time in which the rate exceeds 200 kHz and as the percentage of time in which the rate exceeds 1.2 times the median rate value.
For the data shown in Fig.\ref{fig:rate}, the percentage of bursts larger than 200 kHz is 0.3\% while the percentage of bursts larger than 1.2 times the median rate value is 2.6\%.
In Fig.\ref{fig:rate_time} the PMT rate baseline and the burst fraction as a function of time, calculated averaging in a time interval of 30 min are shown.

\section{Atmospheric muon track reconstruction}

Data recorded during 23$^{rd}$ and 24$^{th}$ January 2007, corresponding to a livetime of 11.3 hours, were analyzed. From the data set analyzed, 2260 atmospheric muon events were reconstructed, corresponding to a reconstruction rate of 0.056 Hz, and their angular distribution was measured \cite{Amore08}.

For comparison, a Monte Carlo simulation of the response of the detector to atmospheric muons was carried out.
A total number of $4\times 10^7$ atmospheric muon
events were simulated with \emph{MuPage} \cite{Carminati08}, corresponding to a livetime
of $4.08\times10^4$ s ($\sim 11.3$ hours). The energy
was sampled between 20 GeV and 500 TeV, while the zenithal angle was
generated in the range $[0^\circ,85^\circ]$.
The real light absorption length spectrum measured at the Test-Site, the optical background evaluated from the measured PMT data rate and the NEMO Phase-1 DAQ electronics read-out, were taken as input for the simulation. The geometry of the detector was simulated averaging, the PMTs positions reconstructed in the whole day in which the data run was acquired, using the acoustic positioning system data.
The Test-Site on-line trigger settings, used during the data acquisition, were also simulated. In order to reject the background hits in the whole sample of hits recorded in an event, off-line trigger and event selection cuts were applied, both in the simulated data and real data analysis. Muon tracks reconstruction was performed using the \emph{reco} code \cite{Aart04}.

\begin{figure}[h]
\begin{center}
  \includegraphics[width=8cm]{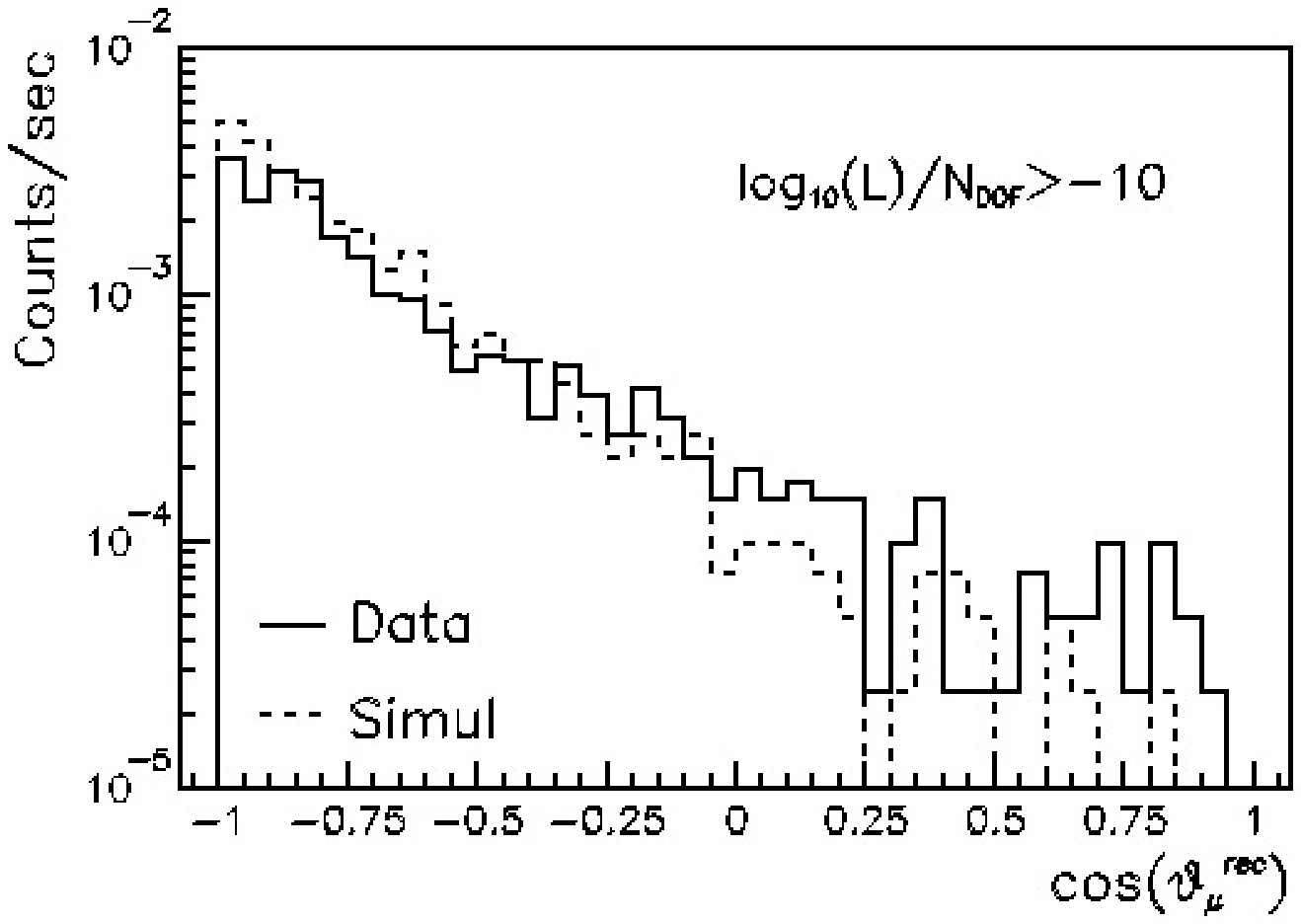}
  \includegraphics[width=8cm]{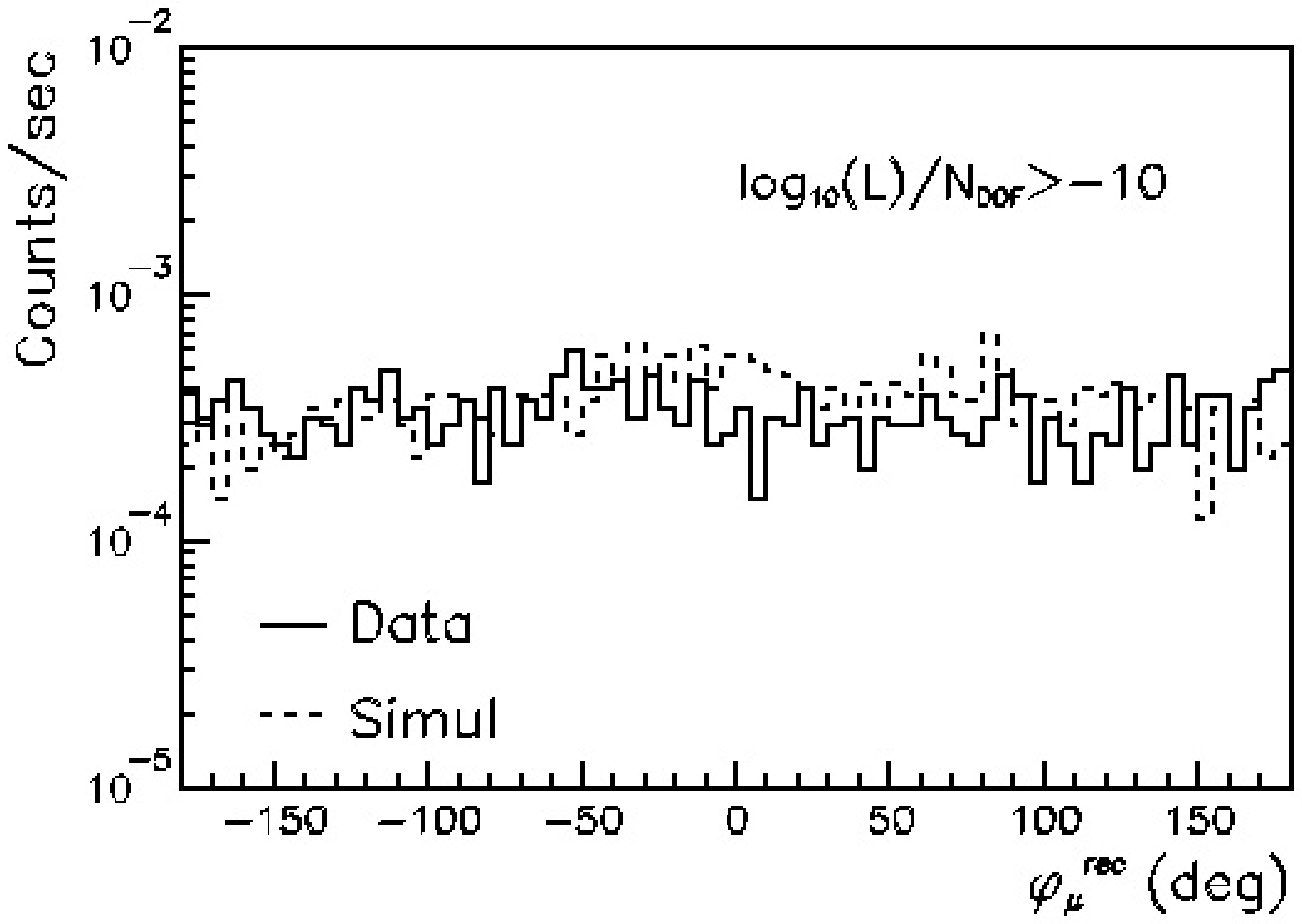}
\end{center}
  \caption{Distribution of reconstructed track directions as a function of the cosine of the Zenith angle ($\cos \theta_{\mu}^{rec}$) (top panel) and of the Azimuth angle ($\phi_{\mu}^{rec}$) (bottom panel). The used likelihood quality cut is equal to -10.}
  \label{fig:angul_distr}
\end{figure}

The angular distributions of reconstructed tracks are shown in Fig.\ref{fig:angul_distr} as a function of the cosine of the Zenith angle ($\cos \theta_{\mu}^{rec}$) and of the Azimuth angle ($\phi_{\mu}^{rec}$).
The data are shown in comparison with simulations. The plotted events were selected applying a quality cut in the likelihood distribution ($L$) for $\log_{10}(L)$/{N$_{DOF}$$>$-10, both for data and simulations.

Despite only a small fraction of the data acquired was analyzed in this work, the comparison
between real data and simulations shows excellent agreement.
The analysis of the whole data set acquired by the NEMO Phase-1 detector (corresponding to a live time of $\sim 3$ months) is in progress.

\section{Conclusions}
The NEMO Phase-1 detector was operated from $18^{th}$ December 2006 to $18^{th}$ May 2007.
The positions of the hydrophones in the tower were calculated with an accuracy
smaller than 10 cm. The magnitude of the underwater current is always smaller than 10 cm/sec and the average direction is $\sim180^{\circ}$. The baseline for each PMT was measured to be $\approx80$ kHz as expected from $^{40}$K decay plus a small contribution of diffuse bioluminescence.
For the data set recorded during 23$^{rd}$ and 24$^{th}$ January 2007, 2260 atmospheric muon events were reconstructed.
The angular distributions of the reconstructed events are compared with simulations and a good agreement is observed.


\begin{thebibliography}{}
\bibitem{Migneco08} E. Migneco et al., \emph{Nucl. Instr. and Meth. in Phys. Res. A} 588 (2008) 111;
\bibitem{Capone08} A. Capone et al., \emph{these proceedings};
\bibitem{Musumeci06} M. Musumeci, \emph{Nucl. Instr. and Meth. in Phys. Res. A} 567 (2006) 545;
\bibitem{Amore08} I. Amore, \emph{Ph.D. thesis}, University of Catania, Italy, (2008)
({\tt http://nemoweb.lns.infn.it});
\bibitem{Carminati08} G. Carminati et al., ``Atmospheric MUons from PArametric formulas: a fast GEnerator for neutrino telescope (MUPAGE)", submitted to \emph{Computer Physics Communications}, (2008);
\bibitem{Aart04}  A. Heijboer, \emph{Ph.D. Thesis}, University of Amsterdam,  The Netherlands (2004)
({\tt http://antares.in2p3.fr/}).
\end{thebibliography}
\end{document}